\documentclass[%
 aip,
 amsmath,amssymb,
 reprint,%
]{revtex4-1}

\usepackage{graphicx}
\usepackage{dcolumn}
\usepackage{bm}
\usepackage{siunitx}
\usepackage{mhchem}
\usepackage{miller}

\usepackage[utf8]{inputenc}
\usepackage[T1]{fontenc}
\usepackage{mathptmx}
\usepackage{etoolbox}

\makeatletter
\def\@email#1#2{%
 \endgroup
 \patchcmd{\titleblock@produce}
  {\frontmatter@RRAPformat}
  {\frontmatter@RRAPformat{\produce@RRAP{*#1\href{mailto:#2}{#2}}}\frontmatter@RRAPformat}
  {}{}
}%
\makeatother

\begin{document}

\title{In-situ azimuthal sample rotation within a horizontal-field cryomagnet for small-angle neutron scattering}
\author{M. T. Littlehales}
\email{matthew.littlehales@tum.de}
\affiliation{Physik-Department, Technische Universität München (TUM), James-Franck-Str. 1, 85748 Garching, Germany}

\author{A. Bauer}
\affiliation{Physik-Department, Technische Universität München (TUM), James-Franck-Str. 1, 85748 Garching, Germany}

\author{M. Bartkowiak}
\affiliation{PSI Center for Neutron and Muon Science, Paul Scherrer Institute (PSI), CH-5232 Villigen, Switzerland}

\author{M. Zolliker}
\affiliation{PSI Center for Neutron and Muon Science, Paul Scherrer Institute (PSI), CH-5232 Villigen, Switzerland}

\author{J. S. White}
\affiliation{PSI Center for Neutron and Muon Science, Paul Scherrer Institute (PSI), CH-5232 Villigen, Switzerland}
 
\author{C. Pfleiderer}
\affiliation{Physik-Department, Technische Universität München (TUM), James-Franck-Str. 1, 85748 Garching, Germany}
\affiliation{Forschungsneutronenquelle Heinz-Maier-Leibnitz (MLZ), Lichtenbergstra\ss e 1, 85747 Garching, Germany}
\affiliation{Munich Center for Quantum Science and Technology (MCQST), Schellingstra\ss e 4, 80799 Munich, Germany}
\affiliation{Zentrum für QuantumEngineering (ZQE), Am Coulombwall 3a, 85748 Garching, Germany}

\date{\today}

\begin{abstract}
We present an in-situ azimuthal rotation stage for small-angle neutron scattering measurements. The stage provides an azimuthal rotation range of \SI{315}{\degree} about the stage normal. At the reference sample-stick orientation, the stage normal is parallel to the incident neutron beam, so that the sample rotates in the plane perpendicular to the beam. Combined with a sample stick rotation about a second, orthogonal axis, the stage provides control of the magnetic-field direction relative to the crystal axes, while the field remains fixed in the laboratory frame. For a \SI{5}{\milli\meter} diameter incident beam, sample-stick rotations of approximately $\pm$\SI{29}{\degree} are accessible with negligible additional attenuation from the holder. The commercial rotator is specified for operation between 1 and 373 K and in magnetic fields up to \SI{35}{\tesla}, making the stage nominally compatible with the full temperature (1.8 to \SI{320}{\kelvin}) and field (0 to \SI{6.8}{\tesla}) ranges of the used cryomagnet. As a working example, we characterize its performance on a spherical single crystal of \ce{MnSi} between 20 and 35 K and in fields up to \SI{183}{\milli\tesla}, observing local sample heating during operation of $\sim$\SI{20}{\milli\kelvin} per \SI{5}{\degree} step. We illustrate the design considerations behind the holder construction, with specific notes for mounting spherical samples, and discuss the resulting performance and limitations. 
\end{abstract}

\maketitle

\section{Introduction}

Neutron scattering and magnetic fields are among the most important experimental tools in condensed matter research. The coupling of a magnetic field to orbital and spin degrees of freedom provides the ability to directly manipulate electron states. Neutron scattering then provides a microscopic probe of the magnetic structures, dynamics, and electron correlations, which in many cases are markedly anisotropic \cite{1985nyephysical}. Examples include orientation-dependent magnetic phase diagrams in rare-earth systems \cite{1991JensenOxf}, quantum phase transitions in Ising magnets \cite{1996BitkoPRL, 2005RonnowScience, 2010ColdeaScience, 2022WendlNature}, exotic magnetic structures \cite{2009MuhlbauerScience, 2012SekiScience, 2018ChaconNatPhys, 2018QianSciAdv, 2023CrisantiPRR}, and anisotropic interactions in chiral helimagnets \cite{2018AdamsPhysRevLett, 2021MoodyPhysRevRes, 2021UkleevPRR, 2023BaralPRR, 2026BaralnpjQuantMatt}. Small-angle neutron scattering (SANS) is the primary method to resolve mesoscale magnetic behavior, where the precise orientation of the magnetic field can be consequential \cite{2022WendlNature, 2018ChaconNatPhys, 2018QianSciAdv}. Typically, SANS measurements use cryomagnets, in which the sample orientation is aligned ex-situ, using a Laue camera or test diffractometer, and fixed during an experiment. Resolving misalignments, or changing the experimental geometry, can consequently waste valuable beam time and limit the feasibility of an experiment. The ability to precisely control the sample alignment or magnetic field orientation in-situ is therefore beneficial. 

Controlling the sample alignment is commonly achieved by dedicated goniometer inserts. For example, inserts have been developed for nuclear magnetic resonance \cite{2012ShirokaRSI, 2019CvitanicRSI, 2021VosegaardNMR}, electronic transport\cite{2014SelbyRSI, 2019WangRSI},  quantum oscillations \cite{2001OhmichiRSI, 2010YeohRSI}, and measurements in pulsed magnetic fields \cite{2020WillisRSI}, all of which utilize the sample rotation to access multiple crystal directions and map angular-dependent physics. These concepts have also been developed for neutron scattering \cite{2010SherlineJPhys, 2014GerberNatPhys, 2016ThalerRSI, 2017LelievreJNR, Gladchenko2019}. However, the angular range is often limited, with the insert developed primarily to correct sample misalignments rather than as an experimental variable. 

Vector magnets provide an alternative that enables precise control of the magnetic field with respect to the sample. These systems use multiple coil sets to control the net field orientation with respect to a fixed sample, and, given their complexity, are often designed for specific use cases \cite{2025DaPratoRevSciInstrum}. Such designs are more constrained for scattering techniques, in which beam access over a wide angular range is required \cite{2014VanderlaanCCR, 2016BarlaJSynchrotronRad, 2017NollVekmag}. Consequently, vector magnets designed explicitly for neutron scattering are rare \cite{ISIS, lavie2016wideaperturevectormagnet}. Furthermore, manipulation of the sample orientation in such systems is typically limited, and so the problems of misalignments or orientation changes persist. 

In this paper, we report the development of an in-situ azimuthal rotation stage for SANS measurements in a horizontal-field cryomagnet. Alongside the sample stick rotation, the stage offers the ability to precisely tune the sample orientation along two orthogonal axes. The combined rotations enable controlled variation of the field direction relative to the crystal axes. This provides an alternative to a vector magnet for measurements within the accessible angular range, with the changing sample orientation accounted for in the data treatment. We subsequently demonstrate this potential by controlling the skyrmion lattice orientation in MnSi during a precession of the magnetic field. Although designed with SANS in mind, we anticipate that our developments may aid in the design of further rotation stages for other neutron scattering techniques where precise control of the sample orientation is either useful or imperative. 

Our paper is organized as follows. In Sec. \ref{Design}, we describe the design of the sample stick and rotation stage along with details of the sample mounting procedure. In Sec. \ref{Characterisation}, we characterize the operation of the rotator stage during an investigation of the skyrmion lattice orientation in \ce{MnSi}. We discuss the maximum scattering angle and resulting background, the consequences of off-center sample mounting, and the thermal performance during operation. Finally, we summarize the main developments and discuss future applications in Sec. \ref{Conclusion}. 

\section{Design and Construction}\label{Design}
The rotation stage was designed to easily integrate with existing sample environment hardware and software, and thus be compatible with any sample stick and cryomagnet fulfilling the requirements outlined below. For the demonstration shown here, we integrated the rotator into the MA7 Oxford Instruments 6.8T horizontal-field cryomagnet at the Swiss Spallation Neutron Source, SINQ, Paul Scherrer Institute (PSI), which has four orthogonal \SI{45}{\degree} access windows that allow the field to be oriented both parallel and perpendicular to the neutron beam without compromise. Together with a variable-temperature insert (VTI), capable of operating from 1.8 to 320 K, the temperature and magnetic field range covers a large parameter space, suitable for a wide range of experiments.

\begin{figure}[b]
	\centering
	\includegraphics{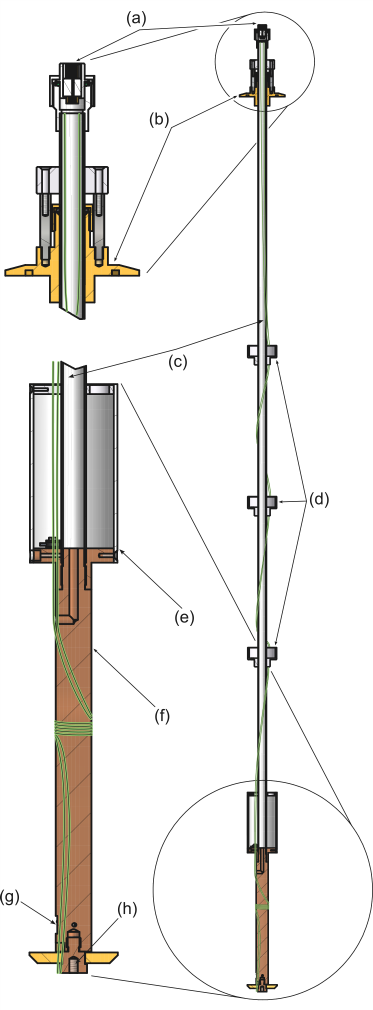}
	\caption{Schematic of the sample stick assembly. (a-h) denote the (a) 10-pin Fischer connector for thermometry and rotator cabling, (b) KF-50 flange to the sample chamber, (c) rotator cabling, (d) outer baffles, (e) aluminum heat exchanger, (f) copper thermal link with integrated heating elements, (g) stick thermometer location, (h) M6 female thread for sample mounting.}
	\label{SampleStick}
\end{figure}

The standard sample stick for MA7 is depicted in Fig. \ref{SampleStick}, consisting of a long stainless-steel tube that terminates outside the sample space in a hermetically sealed 10-pin Fischer connector (a) at the top for cabling and thermometry, a KF-50 flange (b) and radial seal against the tube that seals the stick against the sample chamber and allows for height adjustments, and a copper thermal link at the sample position (f). The sample is cooled by He exchange gas, and three sets of aluminum baffles (d) serve as radiation shields clamped to the stick by a side screw, along with a large aluminum heat exchanger (e). A stick heater is integrated into the thermal link with a thermometer (g) mounted close to the sample position such that the sample stick can be heated independently of the VTI. The stick is also wired for a floating sensor that can be installed close to the sample when necessary for additional temperature sensitivity. Typical samples are mounted to the stick via the M6 thread (h), and so to maintain flexibility, we designed our rotation stage with the same mounting method, avoiding significant modification to the stick. 

\begin{figure}[t]
	\includegraphics{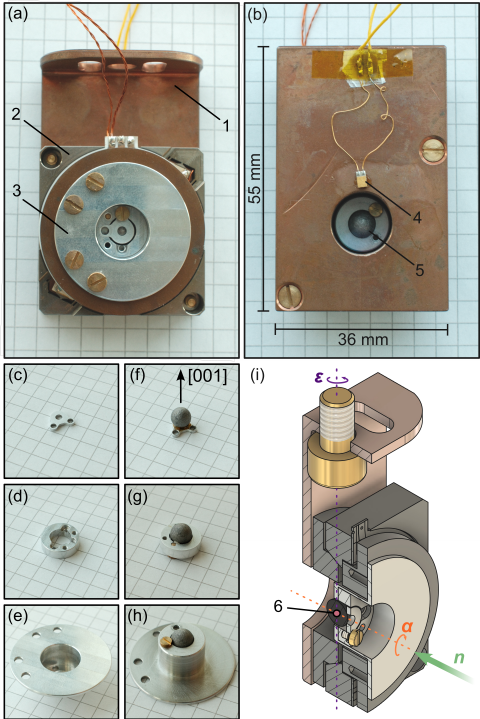}
	\caption{Rotation stage and sample mounting assembly. (a,b) Front and back sides of the sample holder, respectively, consisting of (1) copper mounting plate with (2) ANR240RES (Attocube Systems GmbH, Germany) horizontal-axis rotator, (3) sample mounting assembly, (4) Lake Shore Cernox temperature sensor mounted with GE varnish, and (5) the sample itself. (c-e) Individual elements of the sample mounting assembly, consisting of the sample keyhole mount (c), spacer (d), and mounting bucket (e). All components are fabricated from aluminum. (f) Sample glued onto the keyhole mount with GE varnish, and then mounted into the spacer piece (g). (h) Complete sample assembly with brass screws used to fix all components. (i) Cross-section schematic of the complete sample mount indicating the two rotation axes $\bm{\varepsilon}$ and $\bm{\alpha}$ with dashed purple and orange lines, respectively. The sample coincides with their intersection (6) highlighted by the red dot. The neutron direction is highlighted by the green arrow.}
	\label{RotationStage}
\end{figure}

MA7 contains a large sample-space bore diameter of \SI{48}{\milli\meter}, and is equipped with a rotational unit that allows rotations of the sample stick about its symmetry axis. For the azimuthal rotation component, we chose to use an ANR240RES horizontal-axis rotator (Attocube Systems GmbH, Germany, shown in Fig. \ref{RotationStage}(a)), with a lateral dimension of \SI{35}{\milli\meter}. The ANR240RES is a piezo-controlled rotation stage, constructed from non-magnetic beryllium copper and titanium components, which utilizes a slip-stick mechanism to achieve rotation over an encoded range of \SI{315}{\degree}. The rotator offers relatively low neutron absorption and background scattering, a property aided by the \SI{12.7}{\milli\meter} aperture, and, most notably, offers the ability to mount the sample at the intersection of both the rotator and sample stick rotation axes. 

To maintain compatibility with the standard MA7 sample stick, we designed the holder as shown in Fig. \ref{RotationStage}(a,b), which consists of three main components: the copper mounting bracket (1), the ANR240RES rotator (2), and the sample mounting assembly (3). The copper mounting bracket is an L-bracket of dimensions 36$\times$\SI{55}{\milli\meter} with an aperture milled through the center of the bracket to preserve beam access, as seen clearly from the back profile in Fig. \ref{RotationStage}(b). The ANR240RES mounts to the bracket, and is connected via low-resistance cabling to an ANC350 piezo controller (Attocube Systems GmbH, Germany). The controller is interfaced with frappy \cite{Frappy} which provides easy integration into NICOS or other control systems using SECoP \cite{SECoP, 2020SECoPKiefer}, enabling scriptable control. On the back side of the mounting bracket, we mounted a Lake Shore Cernox thermometer (4), which we refer to as the sample thermometer. Although the thermometer is not in direct thermal contact with the sample (5), we found that this placement yields an accurate measure of the sample temperature when calibrated against the magnetic transition temperature. The thermometer was also used to characterize the thermal performance of the whole rotator assembly during operation and cooling procedures. Both the rotator and thermometer are wired with connector pins and suitable wire slack to allow the complete assembly to be easily removed from the stick when not in use. 

To avoid sample wobble, it is paramount to ensure the sample is maintained in the center of rotation. For measurements in vector fields, it is also important to use regular sample shapes that avoid inhomogeneous demagnetization effects. Spherical samples offer the most homogeneous internal field distribution for different magnetic field orientations. Therefore, we designed a modular mounting assembly with a straightforward mounting process that provides a well-defined rotation axis and maintains the sample at the center of rotation. The assembly consists of three aluminum components fixed together with brass screws as shown in Fig. \ref{RotationStage}(c-h). 

Firstly, the spherical sample is glued to a keyhole piece (c) with GE varnish over a centering hole as shown in (f). To align the sample, we used a Multiwire Laboratories MWL120 back-reflection Laue camera, orienting the [001] crystal direction normal to the keyhole piece (indicated in Fig. \ref{RotationStage}(f)). Precise alignment (within $\sim$\SI{2}{\degree}) along two axes was performed using a manual goniometer stage before the sample was transferred to the keyhole piece. Once the GE varnish had set, the sample alignment was checked and corrected for any misalignments arising during the transfer by softening the GE varnish in isopropyl alcohol, yielding a final accuracy of $\sim$\SI{1}{\degree}. Precise alignment of the sample relative to the keyhole normal is imperative, as any deviations at this stage are propagated through the multiple rotation axes, limiting the accuracy of the field orientation. 

Next, the keyhole piece is screwed into a spacer (d), which is used to align the sample position within the center of rotation of the sample stick. The spacer contains a hole shaped for the keyhole with two orthogonal mounting positions, and is designed in this way to enable two mounting orientations for the sample. Although not necessary for the measurements demonstrated here, this allows the rotation axis to be offset from the aligned crystal direction along two different axes when using a wedged spacer. The spacer also includes a \SI{6}{\milli\meter} aperture for the sample to sit in, as shown by the sample mounted into the spacer in Fig. \ref{RotationStage}(g). These two components are subsequently screwed into the mounting bucket (e) that also includes an aperture for the sample. The complete assembly is shown in Fig. \ref{RotationStage}(h), which then attaches to the rotator such that the sample sits within the rotator body. 

Figure \ref{RotationStage}(i) shows a schematic cross-sectional view of the holder and assembly, with the two rotation axes $\bm{\varepsilon}$ and $\bm{\alpha}$ denoted by purple and orange dashed lines, respectively. As shown, the sample sits directly at the intersection between the two rotation axes, highlighted by a red marker (6). This is achieved by tailoring the dimensions of the spacer specifically to the sample dimensions used here (\SI{5.75}{\milli\meter} sphere). For different sample sizes, the dimensions of the spacer should be tailored accordingly such that the sample is maintained at the center of rotation. Specific dimensions of the sample mounting assembly described here can be found in the Supplementary Material. 

\begin{figure}
	\includegraphics{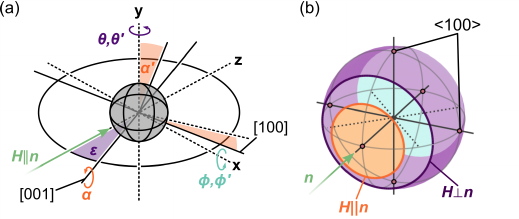}
	\caption{Coordinate system. The SANS geometry includes two rotation axes $\theta$ and $\phi$, which rotate the entire cryomagnet setup. $\epsilon$ denotes the angle of the sample holder normal from the neutron beam direction, and $\alpha$ denotes the azimuthal rotation angle of the sample about the stage normal, such that when $\epsilon = \alpha = $ \SI{0}{\degree}, the \hkl<100> axes are aligned with the laboratory axes, and $\theta', \phi'$, and $\alpha'$ denote the corresponding misalignment angles. (b) Accessible field orientations without scattering through the sample holder. Orange denotes the accessible field orientation, for $\bm{H}\parallel\bm{n}$, and purple for $\bm{H}\perp\bm{n}$, with the red dots denoting the \hkl<100> crystal axes.}
	\label{CoordinateSystem}
\end{figure}

The geometry of an experiment, including the sample rotation, is schematically summarized in Fig. \ref{CoordinateSystem}. The standard magnet tilt angles $\theta$ and $\phi$ are used to obtain rocking scan measurements, where the magnet is rocked through each of these angles by $\pm$\SI{5}{\degree} in steps of \SI{0.25}{\degree} to ensure accurate measurements of the integrated intensities. $\epsilon$ denotes the rotation of the sample stick, where $\epsilon = $ \SI{0}{\degree} orients the sample holder normal (and the sample \hkl[001]) parallel to the neutron beam. The angle $\alpha$  describes the azimuthal rotation of the sample about the rotation-stage normal. At $\epsilon =$ \SI{0}{\degree} and zero magnet rocking angles, this axis i.e. the rotation-stage normal, is parallel to the incident neutron beam, such that $\alpha$ rotates the sample in the plane perpendicular to the beam. Rotation of the sample stick through $\epsilon$ tilts the $\alpha$ axis away from the beam direction; $\alpha$ then describes rotation about this tilted axis. For perfect alignment, at $\epsilon = \alpha = $ \SI{0}{\degree}, the \hkl<100> crystal axes align with the laboratory axes $\bm{x}, \bm{y}, \bm{z}$. Naturally, due to the limited precision of the Laue alignment and sample mounting, there will be some deviation from perfect alignment. The corresponding misalignment angles are denoted by $\theta'$, $\phi'$, and $\alpha'$ in Fig. \ref{CoordinateSystem}(a). 

It is worth emphasizing here that the sample is rotating within the laboratory frame rather than the magnetic field rotating with respect to the sample. This has implications for both analyzing the data and deducing the field direction relative to the crystal axes. When mapping the detector pixels into reciprocal space coordinates, the standard formalism converts the momentum transfer from the laboratory frame $\bm{q}_L$ to reciprocal coordinates $\bm{q}_S$ using a rotation matrix $\bm{R}^{-1}$ such that
\begin{equation}
	\bm{q}_S = \bm{R}^{-1}\bm{q}_L.	
\end{equation}
For standard SANS instruments, the rotation matrix $\bm{R}^{-1}$ includes the $\theta$ and $\phi$ magnet rotations $\bm{R}_\theta^{-1}$ and $\bm{R}_\phi^{-1}$, and $\epsilon$ sample stick rotations $\bm{R}_\epsilon^{-1}$, each defined as follows:
\begin{gather}
	\bm{R}_\theta^{-1} = \begin{pmatrix} \cos\theta & 0 & -\sin\theta \\ 0 & 1 & 0 \\ \sin\theta & 0 & \cos\theta \end{pmatrix}, \\
	\bm{R}_\phi^{-1} = \begin{pmatrix} 1 & 0 & 0 \\ 0 & \cos\phi & -\sin\phi \\ 0 & \sin\phi & \cos\phi \end{pmatrix}, \\
	\bm{R}_\epsilon^{-1} = \begin{pmatrix} \cos\epsilon & 0 & -\sin\epsilon \\ 0 & 1 & 0 \\ \sin\epsilon & 0 & \cos\epsilon \end{pmatrix}.
\end{gather}
The combined rotation matrix is obtained after sequential application of the rotation matrices such that $\bm{R}^{-1} = 	\bm{R}_\theta^{-1}\bm{R}_\phi^{-1}\bm{R}_\epsilon^{-1}$. To include the rotation from our sample stage, we can use an additional matrix, described in the $\epsilon = $ \SI{0}{\degree} condition by
\begin{equation}
\bm{R}_\alpha^{-1} = \begin{pmatrix} \cos\alpha & -\sin\alpha & 0 \\ \sin\alpha & \cos\alpha & 0 \\ 0 & 0 & 1 \end{pmatrix}.
\end{equation}
However, the rotation axis of the sample holder is coupled to the sample stick rotation. Therefore, to account for the $\alpha$ rotation at finite $\epsilon$, a change of basis must be included such that $\bm{U}_\alpha^{-1}$ describes the coupled rotation, where $\bm{U}_\alpha^{-1} = \bm{R}_\epsilon^{-1}\bm{R}_\alpha^{-1}\bm{R}_\epsilon$. The complete transformation matrix from the laboratory frame, including the $\alpha$ rotation axis, is then described by $\bm{q}_S = \bm{U}_\alpha^{-1}\bm{R}^{-1}\bm{q}_L$. Mapping the detector pixels in this way aids in decoupling the extrinsic rotation of the sample from intrinsic effects arising from the changing field direction relative to the crystal axes. For more details of reciprocal space mapping procedures, see Ref. \citenum{2024NevesJApplCrystallogr}. 

The same analysis can also be used to deduce the resulting magnetic field direction relative to the crystal axes. For the sample orientation shown in Fig. \ref{CoordinateSystem}(a), $\bm{R}_\epsilon^{-1}$ and $\bm{R}_\alpha^{-1}$ define the sample stick and sample stage rotations, respectively. To account for the $\theta'$, $\phi'$, and $\alpha'$ misalignments, we can also define the misalignment rotation matrices 
\begin{gather}
	\bm{R}_{\theta'}^{-1} = \begin{pmatrix} \cos\theta' & 0 & -\sin\theta' \\ 0 & 1 & 0 \\ \sin\theta' & 0 & \cos\theta' \end{pmatrix}, \\
	\bm{R}_{\phi'}^{-1} = \begin{pmatrix} 1 & 0 & 0 \\ 0 & \cos\phi' & -\sin\phi' \\ 0 & \sin\phi' & \cos\phi' \end{pmatrix}, \\
	\bm{R}_{\alpha'}^{-1} = \begin{pmatrix} \cos\alpha' & -\sin\alpha' & 0 \\ \sin\alpha' & \cos\alpha' & 0 \\ 0 & 0 & 1 \end{pmatrix}.
\end{gather}
The sample orientation matrix is then given by $\bm{S} = \bm{R}_{\alpha}^{-1}\bm{R}_{\epsilon}^{-1}\bm{R}_{\alpha'}^{-1}\bm{R}_{\phi'}^{-1}\bm{R}_{\theta'}^{-1}$. With the magnetic field $\bm{B}_L$ fixed along the laboratory $\bm{z}$-axis (defined in Fig. \ref{CoordinateSystem}(a)), the magnetic field in the crystal frame is thus $\bm{B}_\mathrm{S} = \bm{S}^T\bm{B}_\mathrm{L}$. Using this methodology, we show the accessible field directions for combined $\epsilon$ and $\alpha$ rotations in Fig. \ref{CoordinateSystem}(b). Accessible directions for the field along the neutron beam are shown in orange, and for the field perpendicular to the neutron beam in purple. This assumes the neutron beam does not pass through the rotator, details of which are given in Sec. \ref{Characterisation}. Notably, for $\bm{H}\parallel\bm{n}$, rotating $\alpha$ when $\epsilon\neq$ \SI{0}{\degree} causes the magnetic field direction in the crystal reference frame to trace a cone about the sample rotation axis. This provides access to field-orientation trajectories that would otherwise require a vector magnet. We will demonstrate this specific case in Sec. \ref{Precession}.

\section{Characterization and Performance}\label{Characterisation}
The rotator is specified by the manufacture to operate at temperatures between 1 and \SI{373}{\kelvin} and under magnetic fields up to \SI{35}{\tesla}\cite{attocube_anr240res}. We tested and characterized the rotation stage only within a temperature range of 20-35 K, and in magnetic fields up to \SI{183}{\milli\tesla}, using the magnetic ordering in a single crystal of \ce{MnSi} as a working example. For all tests, we used operating parameters of \SI{70}{\volt} and \SI{800}{\hertz}, which provided consistent movement without hysteresis across the measured temperature range. The rotator infrequently got stuck during large movements (> \SI{30}{\degree}) owing to heating of the rotator. However, during scripted operation, this was always resolved by sending multiple small rotation commands with \SI{15}{\second} waits to ensure the rotator reached the target setpoint. The commercially provided rotator specifications encompass the nominal operating range of MA7. However, the performance of the complete assembly outside the temperature and field ranges investigated here remains to be characterized, particularly with respect to heating and reliable motion at lower temperatures. 

\begin{figure}
	\includegraphics{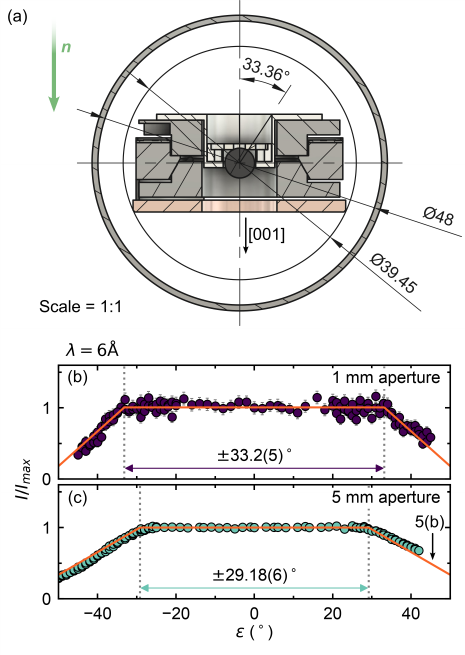}\vspace{-10pt}
	\caption{Scattering calculations and characterization. (a) Schematic cut of the rotator assembly from inside the magnet bore. The sample is maintained in the center of the bore, yielding a maximum opening angle of $\pm$\SI{33.36}{\degree}, assuming an infinitesimal beamspot size. (b,c) Intensity of the transmitted beam normalised to the intensity at $\epsilon =$ \SI{0}{\degree} for neutrons of wavelength $\lambda = $\SI{6}{\angstrom} using a collimation and detector distance of \SI{8}{\meter} and a circular aperture of diameter \SI{1}{\milli\meter} (b) and \SI{5}{\milli\meter} (c). For the \SI{1}{\milli\meter} aperture, the maximum angular range with negligible beam attenuation is $\pm$\SI{33.2(5)}{\degree}, which reduces to $\pm$\SI{29.18(6)}{\degree} for the \SI{5}{\milli\meter} aperture. Dashed lines indicate the maximum window of negligible beam attenuation. The arrow in (c) indicates the point at which the measurements in Fig. \ref{HelixCharacterisation} were performed.}
	\label{ScatteringCharacterisation}
\end{figure}

\subsection{Scattering Angles and Background Characterization}
The aperture in the rotator yields a finite angular range of minimum background and absorption. To commission the rotator stage and characterize the scattering geometry, we used the SANS-I beamline at PSI with a neutron wavelength of $\lambda$ = \SI{6}{\angstrom}, wavelength spread $\Delta\lambda/\lambda$ = 10\%, and detector and collimation distance of \SI{8}{\meter} \cite{2000KohlbrecherJApplCrystallogr}. 

Figure \ref{ScatteringCharacterisation}(a) shows a schematic cross section of the sample holder within the bore of the magnet. Since the mounting system loads the sample within the rotator assembly, the footprint of the holder is kept to a minimum such that it traces a maximum diameter of $\sim$\SI{39.5}{\milli\meter}. This enables full rotation of the complete assembly within the cryostat. With the mounting system described in Sec. \ref{Design}, the angular optical access is $\pm$\SI{33.36}{\degree}. Using a \SI{1}{\milli\meter} circular aperture, we rotated the sample holder within the magnet by an angle $\epsilon$ and measured the transmission of the direct beam. These results are shown in Fig. \ref{ScatteringCharacterisation}(b,c). Fitting the transmission with a trapezoid function (shown by the solid orange line), we find that the angular access reproduces the maximum angle expected from the geometric calculations. We observe a maximum angular access of $\pm$\SI{33.2(5)}{\degree} before the rotator absorption begins to reduce the transmission, yielding the accessible field directions as shown in orange in Fig. \ref{CoordinateSystem}(b). 

Increasing the aperture to a \SI{5}{\milli\meter} diameter reduces the maximum angular access to $\pm$\SI{29.18(6)}{\degree}, as shown in Fig. \ref{ScatteringCharacterisation}(c). Typical SANS measurements involve rocking the complete cryomagnet about the $\theta$ and $\phi$ axes while taking measurements at small angular intervals. The rocking angle and range are dependent on the wavevector, $Q$, and correlation length of the object of interest. The angular access here is still more than sufficient for standard SANS experiments, including commonly studied rare-earth intermetallics that can require $2\theta\approx$\SI{10}{\degree} \cite{2023SinghNatComm}. 

\begin{figure}[t]	\includegraphics{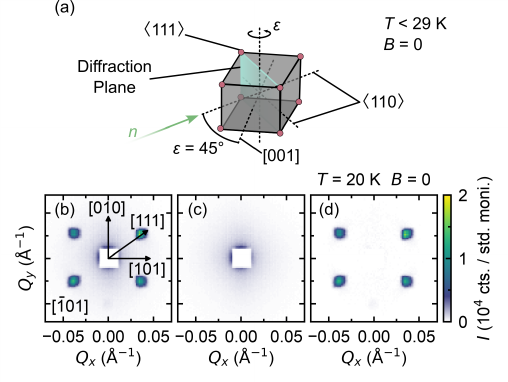}\vspace{-10pt}
	\caption{Sample alignment via helix scattering. (a) Schematic of the sample rotation required to observe the helical magnetic satellites oriented along the \hkl<111> directions. As the sample is oriented with the \hkl[001] direction normal to the rotation stage, the holder must be rotated by \SI{45}{\degree} to reach the diffraction condition for the magnetic satellites. The transmission is consequently reduced owing to scattering through the holder, as indicated in Fig. \ref{ScatteringCharacterisation}(c). (b) Integrated rocking scan from SANS measurements of the helical phase in a \SI{5.75}{\milli\meter} sphere of \ce{MnSi} at $T$ = \SI{20}{\kelvin}. (c) Integrated rocking scan background measurement at \SI{35}{\kelvin} in zero field, measured at identical angular positions. (d) Background subtraction of (b) and (c).}
	\label{HelixCharacterisation}
\end{figure}

For our measurements, we investigated the magnetic ordering in \ce{MnSi}, a well-understood itinerant chiral magnet with the B20 crystal structure. \ce{MnSi} orders magnetically below $T_\mathrm{C}\approx$ \SI{29}{\kelvin}, and the competition between ferromagnetic exchange and the Dzyaloshinskii-Moriya interaction, arising from the lack of crystal inversion symmetry, leads to long-wavelength helical spin textures in the ground state \cite{1976IshikawaSSC}. The periodicity of the helices is approximately \SI{180}{\angstrom}, corresponding to a $Q$-vector of \SI{0.035}{\per\angstrom}, which align along the cubic diagonal \hkl<111> directions owing to leading-order cubic magnetocrystalline anisotropy \cite{2017BauerPRB}. This provides a good confirmation of the sample alignment.  

To observe the helical satellites in the \hkl[001] geometry defined in Sec. \ref{Design}, a rotation in the $\epsilon$ direction is required to reach the diffraction condition. Schematically this is shown in Fig. \ref{HelixCharacterisation}(a). With the mounting geometry and sample alignment described in Fig. \ref{RotationStage}, a sample rotation of $\epsilon = $ \SI{45}{\degree} (denoted by the annotation in Fig. \ref{ScatteringCharacterisation}(c)) is required. At this $\epsilon$ angle, the incident neutron beam passes through the rotator and subsequently scatters through the rotator on the exiting side. From Fig. \ref{ScatteringCharacterisation}(b,c), we would expect a reduction in transmission of approximately half at this angle. 

Full rocking scans, measured over $\pm$\SI{5}{\degree} in steps of \SI{0.25}{\degree}, were taken at this position in $B$ = 0 and $T$ = \SI{20}{\kelvin}, with the integrated rocking scan shown in Fig. \ref{HelixCharacterisation}(b). We also show a background measurement taken at $T$ = \SI{35}{\kelvin} in (c). The background from the sample holder is clearly visible in the raw data, with a maximum count rate of 4.15$\times10^3$ cts. / std. moni..  Owing to the size and scattering strength of the sample, the helical satellites exhibit a count rate of 1.76$\times10^4$ cts. / std. moni., $\sim4.2$ times stronger than the background signal. Therefore, for these measurements, the background from the holder does not hinder wavevector extraction or orientation determination. Nevertheless, for accurate integrated intensities, it is still necessary to perform a background subtraction, as shown in (d). This will be essential for measuring small or weakly scattering samples.

\begin{figure}[t]
	\includegraphics{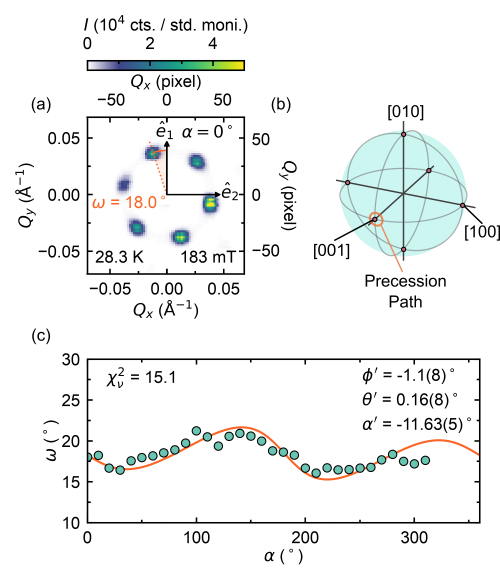}\vspace{-10pt}
	\caption{Demonstration of the sample rotation from the skyrmion lattice in \ce{MnSi}. (a) SANS pattern from skyrmion lattice taken in the center of the equilibrium region at $T$ = \SI{28.3}{\kelvin} and $B$ = \SI{183}{\milli\tesla}, with the precession angle set at $\epsilon = $ \SI{6}{\degree} and $\alpha = $ \SI{0}{\degree}. (b) Schematic of the field precession achieved by offsetting the rotation axis from the field direction. (c) $\omega$ angle versus $\alpha$ when the magnetic field precesses around the \hkl[001] crystal direction. The solid orange line shows the theory of Ref. \citenum{2018AdamsPhysRevLett}, fitting for the misalignment angles defined in Fig. \ref{CoordinateSystem}(a).}
	\label{RotatorDemonstration}
\end{figure}

\subsection{Precession Demonstration}\label{Precession}
To commission the operation of the rotator, we chose to investigate the orientation of the skyrmion lattice in \ce{MnSi}. The skyrmion lattice is an ordered magnetic structure consisting of topologically non-trivial vortex-like whirls of magnetization, stable only within a small region of temperature and magnetic field close to the ordering temperature \cite{2009MuhlbauerScience}. The skyrmion lattice orders with a hexagonal coordination perpendicular to the magnetic field direction, analogous to the flux-line lattice in type-II superconductors, and is defined in reciprocal space by three phase-locked magnetic wavevectors oriented at \SI{120}{\degree} to each other with $Q\sim$\SI{0.039}{\per\angstrom}\cite{2011AdamsPhysRevlett}, as shown in Fig. \ref{RotatorDemonstration}(a). The in-plane orientation of the skyrmion lattice, denoted by $\omega$, is controlled via higher-order magnetocrystalline anisotropy terms. Previous SANS measurements, with a magnetic field rotated through the high-symmetry crystal directions, enabled the sixth-order magnetocrystalline anisotropy parameters to be extracted\cite{2018AdamsPhysRevLett}. These anisotropy parameters fully characterize the in-plane skyrmion lattice orientation for magnetic fields oriented along any crystal direction, and so provide a unique opportunity to test the capability of the rotator. 

To test our setup, we performed a precession of the magnetic field about a high-symmetry \hkl[001] crystal direction where the skyrmion lattice orientation is known to be susceptible to small changes in the field orientation \cite{2018AdamsPhysRevLett}. To generate the skyrmion lattice, we cooled the sample from $T$ = \SI{35}{\kelvin} in a magnetic field of $B$ = \SI{183}{\milli\tesla} to a measurement temperature of $T$ = \SI{28.3}{\kelvin}. By rotating the sample stick to offset the rotation axis from the magnetic field and neutron direction, a rotation of the sample by $\alpha$ creates a precession of the magnetic field within the crystal reference frame, to trace a cone about the \hkl[001] direction. We performed the precession at $\epsilon = $ \SI{6}{\degree}. The resulting projection of the magnetic field along the changing crystal axes is shown schematically in Fig. \ref{RotatorDemonstration}(b) by the orange circle. In the reference frame of the rotating sample, the magnetic field traces a circular precession around the \hkl[001] direction. 

Accounting for the rotating sample frame by mapping the detector pixels into reciprocal space using the equations defined in Sec. \ref{Design}, we can decouple the intrinsic skyrmion lattice orientation from the extrinsic rotating crystal axes. The resulting $\alpha$-dependent skyrmion lattice orientation $\omega$ after reciprocal-space mapping is shown in Fig. \ref{RotatorDemonstration}(c). We included fitting parameters in the theory to characterize any misalignments of the sample. These misalignments naturally arise from the limited precision of the sample orientation with respect to the normal of the keyhole mounting piece as outlined in Sec. \ref{Design}. We determine the offset parameters to be $\phi'$ = \SI{-1.1(8)}{\degree}, $\theta'$ = \SI{0.16(8)}{\degree}, and $\alpha'$ = \SI{-11.63(5)}{\degree}. $\phi'$ and $\theta'$ agree with the tolerances of the Laue alignment discussed in Sec. \ref{Design}, and $\alpha'$ arises simply from a rotational offset between the crystal axes and the rotator zero position, providing a systematic offset of the measured $\omega$ angle.

\subsection{Performance and Limitations}
Our results confirm the functionality of the rotation stage and demonstrate its utility toward a scientific case. It is also worth noting a few experimental details when using the rotation stage. 

\begin{figure}[b]
	\includegraphics{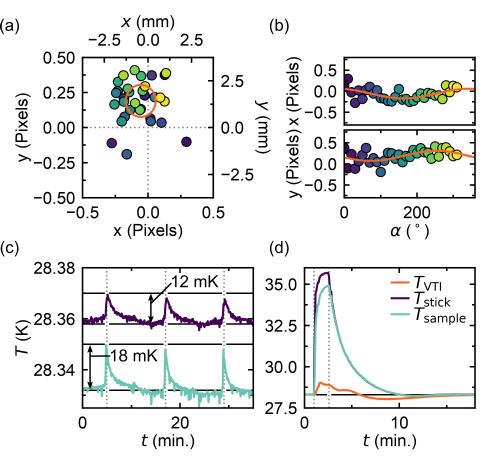}\vspace{-10pt}
	\caption{Rotator diagnostics. (a) Distribution of beam centers calculated from the center position of the skyrmion peaks, assuming a rigid six-fold coordination. (b) Beam center coordinates $x$ (top) and $y$ (bottom) versus sample rotation $\alpha$ with sinusoidal fits demonstrating a slight wobble of the sample. (c) Sample heating from high-voltage actuation of the rotation stage. Purple denotes the temperature measured with the stick thermometer and teal denotes the sample temperature. Rotating in-situ without resetting the magnetic state causes sample heating of $\sim$\SI{18}{\milli\kelvin} at operating temperatures on the order of \SI{30}{\kelvin}. (d) Thermal cycle to reset the magnetic state. The sample is warmed via the stick heater to 35 K in \SI{90}{\second} before the heater is switched off and the sample cools to \SI{28.3}{\kelvin} over a duration of 15 minutes.}
	\label{TempCharacterisation}
\end{figure}

Although great care was taken in maintaining the sample within the center of rotation of both the sample stick and the rotator axis itself, it is inevitable that small offsets, and the stacking of tolerances, will lead to a slight deviation from the center of rotation. This causes a slight sample wobble that leads to a moving beam center in the scattering pattern, which can be a major source of error in extracting positions and magnitudes of wavevectors from SANS measurements. In Fig. \ref{TempCharacterisation}(a), we show the distribution of the beam centers as determined by considering the scattering pattern as a rigid six-fold structure \cite{2018AdamsPhysRevLett} and determining the intersection between the six wavevectors. Fig. \ref{TempCharacterisation}(b) then shows the $\alpha$ evolution of the $x$ and $y$ coordinates. The good fit of a sinusoid to the $x$ and $y$ positions of the beam center, shown by the orange line in (b), suggests that the sample is wobbling, likely caused by an off-center mounting position. By determining the amplitude of the wobble, and propagating it back to the sample position, we find that the off-center distance $d = $ \SI{0.36(7)}{\milli\meter}. This is small, given the stacked machining tolerances from the components of the sample mounting assembly. The resulting beam centers for each $\alpha$ position are distributed over a range less than the width of a full detector pixel. However, the high $Q$-resolution in SANS may still be affected by a moving beam center, artificially indicating a changing or asymmetric wavevector. This highlights the necessity to ensure that the sample is close to the center of rotation, and of considering the moving beam center on a case-by-case basis in the data analysis.

The sample thermometer was used to characterize the thermal response of the holder during standard cooling procedures and during rotation. Despite the large mass inside the VTI, the system responds very well, with negligible differences observed between the sample and stick temperatures in the 20-\SI{35}{\kelvin} range studied here. In Fig. \ref{TempCharacterisation}(c) and (d), we show the thermal response of the rotation in two distinct cases. 

Firstly, in (c), we show the temperature measured on the stick (purple) and at the sample (teal) over a 35-minute period during which the rotator position was changed three times without changing the temperature of the VTI (maintained at \SI{28.3}{\kelvin}). The small constant offset of the sample temperature from the VTI bath arises from the PID settings on the cryomagnet, and the dashed lines indicate the times at which the rotation position is changed. The slip-stick mechanism of the piezo rotator utilizes small amounts of friction to achieve the large rotation angles while maintaining high angular resolution. Accordingly, a \SI{5}{\degree} rotation leads to a small temperature increase of \SI{18}{\milli\kelvin} at the sample position with a \SI{12}{\milli\kelvin} increase at the stick thermometer, and an equilibration time of approximately 15 minutes. The difference in the two temperature readings suggests local heating of the rotator, and therefore the sample. On a background of \SI{28.3}{\kelvin}, this temperature increase is negligible. However, for lower measurement temperatures, this may become significant\cite{2014GerberNatPhys}. Furthermore, the sticking observed for large rotation angles may worsen for lower temperatures. For large rotation angles, smaller incremental rotations followed by regular waiting periods should be used to circumvent considerable heating and sticking. Alternatively, it may be necessary to employ a temperature-cycling process. 

During temperature cycling, the sample temperature is increased above the characteristic ordering temperature before the rotation is performed, such that local heating during rotation does not affect the measurement. In (d), we demonstrate the thermal response of this procedure, with the orange, purple, and teal solid lines indicating the VTI, stick, and sample temperatures, respectively. The VTI setpoint is maintained at the measurement temperature (\SI{28.3}{\kelvin}) throughout the duration of the measurement. Once a measurement is complete, the stick heater is switched on (indicated by the first gray line), and the sample takes roughly \SI{90}{\second} to reach the setpoint of \SI{35}{\kelvin}. During heating, there is a slight delay between the stick and sample temperatures. However, once the sample reaches the setpoint, the rotator position is changed, and the heater is switched off (indicated by the second gray dashed line). The stick and sample temperatures then closely track each other on cooling, indicating that there are negligible heat gradients across the stick and rotation stage within this temperature range. The sample and VTI then take roughly 15 minutes to equilibrate, with the VTI temperature changing by approximately \SI{1}{\kelvin} during the complete temperature cycle. 

For the example measurements shown in Sec. \ref{Precession}, the thermal resetting procedure was imperative, owing to pinning effects restricting the skyrmion lattice reorientation \cite{2018AdamsPhysRevLett}. Accordingly, this resetting process restricts the measurement frequency. For this sample and measurement protocol, we achieved a measurement duty cycle of approximately 50\%, consisting of rocking-scan measurements over a duration of 15 minutes, followed by the 15-minute thermal reset procedure indicated above. However, without a thermal reset, the measurement frequency is not limited by the sample rotation. 

Among the limitations arising from neutron background, heating (relevant for low temperatures), and sticking for large rotations, the main limitation of the current design is the finite range of the resistive encoder. Although it is possible to drive the rotator indefinitely, driving it beyond the encoded range may lead to a loss of position requiring realignment ex-situ. Integrating a homing switch and calibrating the rotator optically may be a possible solution if full \SI{360}{\degree} operation is required. Furthermore, masking the periphery of the rotator with a strong neutron absorber such as cadmium and replacing the brass screws with aluminum may aid in reducing the background for measurements of weakly scattering samples, but would limit any measurements requiring larger scattering angles. 

\section{Conclusions}\label{Conclusion}
In conclusion, we report on the development of a stage suitable for in-situ two-axis sample rotation for SANS that, per manufacturer specification, is compatible with temperatures down to \SI{1}{\kelvin} and magnetic fields up to \SI{35}{\tesla}. Using a single crystal of \ce{MnSi} between 20 and \SI{35}{\kelvin} and up to \SI{183}{\milli\tesla}, we demonstrate a simple mounting method for spherical single crystals that enables precise control of the magnetic field with respect to the crystal axes, using the field-direction-dependent skyrmion lattice orientation during a magnetic field precession to highlight its potential. 

We anticipate that the rotation stage assembly shown here may be useful for a number of experiments in which precise control of the field orientation with respect to the crystal axes is important. For example, in-situ rotation may be useful to perform angle-dependent scattering measurements to aid in distinguishing single-$Q$ versus multi-$Q$ spin textures \cite{}. Furthermore, fine control of the relative field orientation may be useful to explore transverse Ising systems where mesoscale quantum phase transitions are observed as the magnetic field is tilted away from the hard axis \cite{2022WendlNature}. The combination of the fixed magnetic field and two-axis sample rotation provides an alternative to a vector magnet for controlling the field direction relative to the crystal axes over the accessible angular range. In principle, the rotator is not limited to SANS measurements and may be suitable for wide-angle diffraction measurements. However, the powder patterns and background arising from the rotator and assembly will be considerably larger, and absorption corrections may be necessary. Nevertheless, we anticipate that these considerations may be useful for the development of more specialized holders for precise alignment and vector-field capabilities in other neutron scattering techniques. 

Our design specifically addresses the straightforward mounting of spherical samples. However, with this paper, we aim to provide a platform for future studies that may benefit from in-situ sample reorientation or precise control of the magnetic field orientation. Therefore, we expect that with minor adaptations to the design, and taking note of the mounting considerations discussed here, it will be possible to mount samples of other geometries where spherical samples may be unobtainable. 

\section*{Acknowledgments}
We wish to thank M. Pfaller and the team of the Zentralwerkstatt of the Physik-Department at TUM. We gratefully acknowledge discussions with M. Garst and technical support from S. Giemsa and S. Stamm. This work is based on experiments performed at the Swiss Spallation Neutron Source SINQ, Paul Scherrer Institute, Villigen, Switzerland, under proposal number 20260123. This study received funding from the Deutsche Forschungsgemeinschaft (DFG, German Research Foundation) under TRR80 (From Electronic Correlations to Functionality, Project No.\ 107745057), TRR360 (Constrained Quantum Matter, Project No.\ 492547816, Projects A6, C1, and C3), SPP2137 (Skyrmionics, Project No.\ 403191981, Grant PF393/19), and the excellence cluster MCQST under Germany's Excellence Strategy EXC-2111 (Project No.\ 390814868). Financial support by the Bundesministerium für Forschung, Technologie und Raumfahrt (BMFTR, Federal Ministry of Research, Technology and Space) through the project SIM-Mode (Project No. 05K22WO2) and by the European Research Council (ERC) through Advanced Grants No.\ 291079 (TOPFIT) and No.\ 788031 (ExQuiSid) is gratefully acknowledged.

\section*{Author Declarations}
\subsection*{Conflict of Interest}
\noindent The authors have no conflicts to disclose.
\subsection*{Author Contributions}
\noindent\textbf{Matthew Littlehales}: Conceptualization (equal); Data curation (lead); Formal Analysis (lead); Investigation (equal); Methodology (equal); Software (equal); Visualization (lead); Writing - original draft (lead). \textbf{Andreas Bauer}: Conceptualization (equal); Resources (equal); Writing - review \& editing (equal). \textbf{Marek Bartkowiak}: Resources (equal); Writing - review \& editing (equal). \textbf{Markus Zolliker}: Software (equal); Writing - review \& editing (equal). \textbf{Jonathan White}: Investigation (equal); Writing - review \& editing (equal). \textbf{Christian Pfleiderer}: Conceptualization (equal); Supervision (lead); Writing - review \& editing (equal).

\subsection*{Data Availability}
\noindent The data that support the findings of this study are available from the corresponding author upon reasonable request.

\section*{References}
\bibliography{bib}

\end{document}